\documentclass[12pt,a4paper]{article}
\usepackage{jheppub}

\usepackage{graphicx}
\usepackage{subcaption}
\usepackage{xspace}

\usepackage[utf8]{inputenc}

\usepackage[style=numeric-comp,sorting=none]{biblatex}
\abstract{The High Energy Physics community has developed and needs to maintain
many tens of millions of lines of code and to integrate effectively the
work of thousands of developers across large collaborations. 
Software needs to be built, validated, and deployed across hundreds of
sites. Software also has a lifetime of many years, frequently beyond that
of the original developer, it must be developed with sustainability in
mind. Adequate recognition of software development as a critical task in the HEP
community needs to be fostered and an appropriate publication and citation
strategy needs to be developed. As part of the HEP Software Foundation's
Community White Paper process a working group on Software Development,
Deployment and Validation was formed to examine all of these
issues, identify best practice and to formulate recommendations for the next
decade. Its report is presented here.}

\begin{document}

\noindent
\begin{tabular*}{\linewidth}{lc@{\extracolsep{\fill}}r@{\extracolsep{0pt}}}
 & & HSF-CWP-2017-13 \\
 & & December 21, 2017 \\ 
 & & \\
\end{tabular*}
\vspace{2.0cm}

\title{HEP Software Foundation Community White Paper Working Group -- Software
Development, Deployment and Validation}

\author{HEP Software Foundation:}
\author[a]{Benjamin Couturier,}
\author[a,1]{Giulio Eulisse,}
\author[b]{Hadrien Grasland,}
\author[a]{Benedikt Hegner,}
\author[b]{Michel Jouvin,}
\author[c]{Meghan Kane,}
\author[d]{Daniel S. Katz,}
\author[e]{Thomas Kuhr,}
\author[f]{David Lange,}
\author[a,1]{Patricia Mendez Lorenzo,}
\author[e]{Martin Ritter,}
\author[a,1]{Graeme Andrew Stewart,}
\author[a]{and Andrea Valassi}

\affiliation[a]{CERN, Geneva, Switzerland}
\affiliation[b]{LAL, Université Paris-Sud and CNRS/IN2P3, Orsay, France}
\affiliation[c]{Soundcloud, Berlin, Germany}
\affiliation[d]{University of Illinois Urbana-Champaign, Champaign, Illinois,
USA}
\affiliation[e]{Fakultät für Physik, Ludwig-Maximilians-Universität München,
München, Germany}
\affiliation[f]{Princeton University, Princeton, USA}
\affiliation[1]{Paper Editor}

\maketitle

\newpage

\hypertarget{scope}{%
\section{Scope}\label{scope}}

As part of the process to layout the HEP Software Foundation's Community White
Paper Roadmap~\cite{HSF-CWP-2017-01} we formed a group to study issues around
software development, deployment and validation. The starting point for the
group was to recognise that modern High Energy Physics experiments are large
distributed collaborations comprising up to a few hundred people actively writing
software. It is therefore central that the processes and tools used for
development are streamlined to ease the task of contributing code and
to facilitate collaboration between geographically separated peers. At
the same time we must properly manage the whole project, ensuring code
quality, reproducibility and maintainability with the least effort
possible. Making sure this happens is largely a continuous process, and
shares a lot with non-HEP specific software industries.

The group is therefore interested in tracking and promoting solutions
for the following areas:

\begin{itemize}
\item
    Distributed development of software components, including the tools
  and processes required to do so (code organisation, documentation,
  issue tracking, artifact building) and the best practices in terms of
  code and people management.
  \item
    Software quality, including aspects such as modularity and reusability
  of the developed components, sustainability of the development effort,
  architectural and performance best practices.
  \item
    Deployment of software and interaction with operations teams.
  \item
    Validation of the software both at small scales (e.g., best practices
  on how to write a unit test) and larger ones (large scale validation
  of data produced by an experiment).
  \item
    Software licensing and distribution, including their impact on
  software interoperability.
  \item
    Recognition of the significant contribution that software makes to
  High Energy Physics as a field.
\end{itemize}

The goal of this document is to highlight the challenges the field will
face in the next 5 to 10 years, and to provide a strategy on how address
them. Adopting best practices from non-HEP communities and adapting them
to HEP-specific needs will be key for the success of the effort.

\hypertarget{main-challenges-in-the-next-5-10-years}{%
\section{Main challenges in the next 5-10
years}\label{main-challenges-in-the-next-5-10-years}}

The challenges can be divided into two broad categories: those generic
to large-scale distributed software development and those specific to
the HEP environment.

The HEP-specific challenges are those derived from the fact that HEP is
a large, inhomogeneous community with multiple sources of funding,
mostly formed of people belonging to small university groups and some
larger laboratories. Software development effort within an experiment
usually encompasses a huge range of experience and skills, from a few
more or less full time experts to many physicist programmers with little
formal software training. In addition, the community is split between
different experiments that often diverge in timescales, size and
resources.

Another peculiarity of HEP is that the experiment software is
usually divided in two separate use cases, production (data acquisition,
data reconstruction or simulation) and user analysis, whose requirements
and life-cycles are completely different. The former is very carefully
managed in a central and slow moving manner, following the schedule of the
experiment itself. The latter is much more dynamic and strongly coupled
with conferences or article publication timelines. Finding solutions
which adapt well to both cases is not always obvious or even possible.

Principle areas of challenge are:

\paragraph{Effective use of developer effort}

\begin{itemize}
\item
    The large fraction of short term students and collaborators
  found in our field requires a process for evaluating software
  contributions to common projects and for assuring not just their
  quality, but also their maintainability.
  \item
    For the same reason, we need to minimise the time required for people
  to begin to contribute effectively.
\end{itemize}

\paragraph{Knowledge sharing}

\begin{itemize}
\item
    There is a strong need for knowledge sharing, including effective
  training. This is required by groups that are geographically
  separated, but also amongst different fields, with different skills,
  languages and cultures. E.g., some students might be able to track new
  trends better than older colleagues, but they might lack vision;
  software engineers might provide better implementations than
  postdocs, but may lack a complete understanding of the broader
  picture for the experiment needed to set priorities.
  \item
    There is an historical and stubborn compartmentalisation between
  different experiments, so propagating knowledge horizontally across
  experiments is particularly hard.
\end{itemize}

\paragraph{Reusing code effectively}

\begin{itemize}
\item
    While striving towards common solutions, the group recognises that a
  single platform or set of tools will not fit all use cases, but we
  emphasise that generic best practices and recommendations can be made.
  A top-down approach, where a given tool or solution is mandated has
  not proven to be viable in the past. However, if we can
  identify success stories and make sure they can be extended (if
  needed) and advertised as possible general solutions, 
  duplicate efforts or developments are more likely to be avoided.
\end{itemize}

\paragraph{Validation of complex output}

\begin{itemize}
\item
    Efficient validation of the complex outputs of HEP code, involving
  inputs from highly specialised experts and unique sources, is hard to
  do, especially taking into account stochastic processes and long
  tails.
\end{itemize}

\paragraph{Data, code and knowledge preservation}

\begin{itemize}
\item
    Long term maintainability of the code to handle data that was obtained
  decades before as required by the lifetime of a HEP experiment.
\end{itemize}

\paragraph{Code lifecycle and deprecation}

\begin{itemize}
\item
    In a highly demanding software environment such as the HEP
  communities, there is great pressure to adopt updated versions of
  tools and libraries for performance or design benefits. However, there
  is little corresponding pressure to deprecate old components that may
  have been used for years, causing a bloat of the active software stack
  and provoking crises when abandoned upstream libraries start to fail
  to build. Hence there is a need to engage in effective, smooth and
  continuous lifecycle management of the codebase.
\end{itemize}

\paragraph{Hardware and software evolution}

\begin{itemize}
\item
    Hardware trends are now towards increasing processing capacity through
  concurrency (many cores and vectorisation) rather than any increase in
  serial processing capacity. This is at odds with the traditional HEP
  software workflows, which have largely been serial. Memory latencies
  and a deep memory and storage hierarchy have become key drivers of
  performance. Increasingly specialised processors are also arriving in
  the marketplace, which are not naturally suited to HEP codes. Adapting
  to this new hardware environment and maintaining the efficiency of HEP
  code becomes increasingly hard.
  \item
    New innovations in software development usually come from outside of
  the HEP community. These can provide significant advantages in
  exploiting new hardware or aiding efficient distributed development
  and such technologies need to be tracked, investigated and deployed.
\end{itemize}

\paragraph{Software deployment}

\begin{itemize}
\item
    As HEP software today runs in widely heterogeneous environments, from
  traditional grid computing sites to High Performance Computing sites,
  including the gamut of academic and commercial cloud services, there
  is a challenge to uniformly provide the appropriate software in all of
  these environments. End user analysis code also needs to be managed
  and distributed into these environments. CVMFS~\cite{7310920} has been hugely
  successful here, however it does not cover all cases and a viable
  common solution may emerge and be adopted more widely.
\end{itemize}

\paragraph{Recognition of software development}

\begin{itemize}
\item
    Writing high quality software is a skilled activity that greatly
  contributes to the success of the field. Ensuring that software
  developers receive recognition for their contributions will help
  retain people in the field and help their career development.
\end{itemize}

\hypertarget{meeting-the-challenges}{%
\section{Meeting the Challenges}\label{meeting-the-challenges}}

The previous section introduced a brief summary of the main challenges
this working group has identified. In the next paragraphs we both
expand on these and identify viable existing solutions and areas where
more investigation is required.

\hypertarget{effective-use-of-developer-effort}{%
\subsection{Effective use of developer
effort}\label{effective-use-of-developer-effort}}

\subsubsection{Design documentation and training}

For developers to be able to contribute effectively to experiment
software it is first of all necessary for them to have a clear idea of
how to write code that will mesh effectively with the framework or the
rest of the software stack. Thus it is essential that developers can
access clear documentation, with good examples as to how to model their
code. Effective contact with experts should be available so that good
design choices are made from the outset.

Once a clear picture of the framework is available to new developers the
first task of the contributor should be to make their intentions clear
to others. A common technique is to use a request for comments document
which outlines the design and allows other developers to provide
feedback before substantial coding effort is expended. Such documents
can be accompanied by demonstrator or prototype code to better
illustrate the purpose of the of the changes. This approach allows
documentation and testing to be incorporated into the development itself
and not left as an afterthought. This model was used by the Concurrency
Forum~\cite{ConcurrencyForum} in the past
and was very successful in selecting the most promising approaches for
the community and disseminating knowledge amongst different experiments.

The task for developers who might contribute a general purpose piece of
code that is used across experiments is more challenging as active
solicitation of input will need to be sought. However, if experiment
documentation is open to the HEP community then different experimental
boundary conditions will be more easily understood and translated into
the design and interface of a software product.

In general, good software design principles should be followed and thus
generic training in software design, but with a focus on the HEP problem
space, is a valuable investment for the community. We note that several
computing schools exist already, but that overall only a small fraction of
HEP students attend them as they are considered quite specialist.
Universities, in both their undergraduate and graduate programmes, play an
important role too in core skills training (and in some institutes also
in relevant advanced topics), but this can be uneven. Effort should be
made to incorporate more generic training into the education programmes of
the experiments; centering these activities around some of the larger
labs will also allow sharing of the training effort across different
experiments.

\hypertarget{standard-development-tools}{%
\subsubsection{Standard development
tools}\label{standard-development-tools}}

To maximise the productivity of developers it is very beneficial to use
development tools that are not HEP-specific. There are many open-source
projects and tools that are of similar scale to large experiment
software stacks and standard tools are usually well documented. Many
simple questions can be answered using standard resources, such as Stack
Overflow.

Insofar as an experiment requires customisation of tools, documentation
and training are evidently important. However, the costs of such
customisation can be easily outweighed by the cost of maintenance and
training, as well as a possible loss of generic skills for users.

For source control in particular, the open-source world has moved
strongly towards distributed version control as superior to centralised
solutions. The advantages for the individual developer are numerous,
including independence from central services and greater freedom to
experiment. Finished code enhancements can be proposed using a
pull request workflow so that high quality contributions are made with
the central repository receiving pristine patches. We note that the
community has generally chosen to move to git. This standardisation
is very welcome as it also brings an alignment with many open-source
projects and commercial organisations.

The HEP current community is widely using CMake for the builds of
software packages. This is an open-source system dedicated to building,
testing, and installing software that captures the specific build
specifications for each particular package in a platform independent
way. It is highly regarded because of its modular and scalable
characteristics and is also very widely used outside HEP with a lot of
support available. CMake's use of plugin modules allows scope for
sharing between HEP experiments in the cases of building HEP specific
code, though at the moment probably the best use is not yet made of this
feature in our community. CMake can orchestrate the builds of large
numbers of packages (e.g., the LCG releases of over 300 packages), but
can become quite specific when doing so and thus more limited. The HSF
Packaging group found
Spack~\cite{HSF-TN-2016-03,Gamblin:2015:SPM:2807591.2807623}
to be a promising
more general solution for the task of building an HEP software stack, however, it
currently has a relatively small user base, centered on HPC, and might
just become a niche product with little traction in the wider
development community. More generic tools, such as Homebrew or
Conda are interesting but may not have all the features needed by
HEP experiments.
Thus this area still requires close attention to active developments in
the open-source world.

For reporting problems, tracking bug fixes and planning improvements to
the software issue trackers should be used. The best of these offer
multi-level grouping of issues and provide direct support for useful
aspects of agile development discussed below. Optimal choices will offer
good integration with source code management (e.g., GitHub Issue
Tracker, Jira, GitLab Issue
Tracker) and the choice should usually be made in parallel
with an source code management tool.

\hypertarget{development-environment}{%
\subsubsection{Development environment}\label{development-environment}}

It is important that developers can focus on the design and
implementation of the code and do not have to spend a lot of time on
technical issues. Clear procedures and policies must exist to perform
administrative tasks in an easy and quick way. This starts with the
setup of the development environment, for example by having a minimal
set of requirements on the operating system and the installed packages.
Supporting different platforms not only allows the developers to use
their machine directly for the development, it also provides a check of
portability basically for free.

Other typical workflows are the update of versions or the submission of
code for review. It decreases the turnaround time and reduces
frustration if code quality criteria and policies are checked as early
as possible, ideally on commits to the local repository or even by the
editor. Modern IDEs, which offer good integration with modern
development tools, can be particularly valuable for novice coders. They
provide support for developers in navigating the code base as well as
catching elementary errors. It would be of great benefit for the
community to ensure that standard project templates
(e.g., HSF Project Template~\cite{HSFProjectTemplate}) work
with minimal configuration with standard IDEs, such as Eclipse, XCode,
KDevelop, Qt Creator, etc. For large experiment code bases, the scaling
of these IDEs to many millions of lines requires investigation.

Finally, support is important. A system that enables developers to
get quick and competent advice from experts will improve their
productivity.

\hypertarget{social-coding-sites}{%
\subsubsection{Social coding sites}\label{social-coding-sites}}

The advantages of social coding sites are that they allow developers to
share and discuss code through very streamlined and highly functional
web interfaces. Distributed version control systems are key enablers for
these sites. There are commercial choices, such as GitHub or BitBucket,
that can serve a community well; alternatively a privately hosted
service, such as CERN's GitLab instance, may mesh more easily with other
related services. We note that ``bridging'' code between different service
platforms has been developed by a number of experiments and could be shared
across the community. Smaller projects would probably
benefit from the increased exposure and larger development community from being hosted on GitHub.
For attribution and credit, GitHub also offers a more standard platform
that can be helpful to a developer's career. However, the key
requirement here is the ability to discuss and revise code changes in a
recorded fashion as they become part of the code repository.

\hypertarget{continuous-integration}{%
\subsubsection{Continuous integration}\label{continuous-integration}}

Proper testing of changes to code should always be done in advance of a
request being accepted. Continuous integration, where merge or pull
requests are built and tested in advance is now standard practice in the
open-source community and in industry. HEP software should not be
different. Not only should continuous integration run unit and
integration tests, it can also incorporate code quality checks and
policy checks that will help improve the consistency and quality of code
at low human cost.

For running continuous integration and build orchestration many options
exist (Jenkins, Gitlab CI, Travis, Bamboo) and the particular choice of
which one to use is a relatively pragmatic decision, with only a very
limited impact on the ability to cooperate around software.

\hypertarget{code-review}{%
\subsubsection{Code review}\label{code-review}}

Incorporation of code into the code base should usually also involve
sign off from at least one relevant expert who can check for
architectural violations, anti-patterns or even just simple coding
errors. Code review should be seen as a process of dialogue between the
developer and the reviewer (community expert) and it should be firm, but
friendly.

As well as a way to validate code, reviews can also be a way for
newcomers to learn about the architecture of a project. As part of a
mixed experience reviewing team, fresh eyes on a piece of software can
produce original comments, but the discussions occurring as a side
effect of the review effectively helps new developers understand the
issues at stake.

\hypertarget{agile-development}{%
\subsubsection{Agile development}\label{agile-development}}

HEP is a community where the developer of a piece of software is also
likely to be the `customer' as well. However, despite this, many useful
ideas can be taken from the various agile methodologies, in particular
with regard to solving only the problem at hand and keeping development
cycles as short as possible to find iterative and evolutionary
solutions. Agile development methodologies for physicists, and how to
select the most appropriate one, might be best taught by example rather
than as formalism.

One aspect of agile methodology that would serve HEP rather well is the
concept of a user story that describes the purpose of a system, rather
than requirements capture that was used in the past. Such a method is
more flexible in adapting to changing requirements, which can be quite a
common situation in HEP, especially in the analysis domain.

HEP would benefit from adopting agile retrospectives. A retrospective is
a formalised way for contributors to meet regularly to discuss ways to
improve their software development process. Since a retrospective occurs
regularly, there is a chance to catch issues affecting the development
process as they arise. Issues can be tracked and mitigated over time
rather than ballooning into daunting barriers. Additionally, the focus
is on learning from failures quickly, without placing blame. This
welcomes feedback, positive or negative, toward the goal of improving
the software development process.

Retrospectives can be structured as needed to suit HEP, but the standard
approach lends itself nicely to solving current problems since the last
retrospective:

\begin{itemize}
\item
    Gather data. What went well? What didn't go well? What needs to be
  improved? Each person who has input should have a chance to succinctly
  state their observations.
  \item
    Discuss top issues in prioritised order. Based on the feedback of the
  group, discuss top issues to reach an understanding about the root
  causes.
  \item
    Assign action points. Based on the pain points discussed, assign
  action items that are small enough in scope that they can be resolved
  before the next retrospective.
\end{itemize}

This setting of information sharing and collaboration can yield clear
productivity benefits. For example, it is possible that a developer
experiences issues that slow down their productivity without being aware
of it. Perhaps the developer became accustomed to the build time being
very high and thinks of it is as given constraint. In a retrospective, a
discussion about build times may arise along with known solutions for
reducing the build times. This will elicit awareness about the issue, a
solution, and a subsequent gain in productivity.

Agile development is very well supported by the modern development
infrastructure outlined above.~

\hypertarget{knowledge-sharing}{%
\subsection{Knowledge sharing}\label{knowledge-sharing}}

\hypertarget{documentation}{%
\subsubsection{Documentation}\label{documentation}}

As was mentioned above, proper training and documentation is key to
efficacious use of developer effort. For documentation that has to be
specific, favoured solutions would have a low barrier of entry for
contributors, but also allow and encourage the review of material.
Consequently it is very useful to host documentation sources in a
repository with a similar workflow to code and to use an engine that
translates the sources into modern web pages. An example would be using
the Jekyll engine, as the main HSF website currently does, or using
services such as Read The Docs.

Traditionally HEP has often used Doxygen for reference
documentation, especially given its emphasis on C++. New documentation tools are also
available that integrate Doxygen with better solutions in other
languages, e.g., the python Sphinx documentation tool can use
Breathe to integrate C++ and python documentation. Sites such as
Read The Docs also take much of the pain out of producing an attractive documentation site.

\hypertarget{training}{%
\subsubsection{Training}\label{training}}

Code development, as a critical and transferable skill, is well
recognised inside the HEP software community. Various schools exist that
train students in good coding practice, but only a small minority of
contributors to HEP code attend these. There is therefore scope for
broader and more accessible training that experiments can play a role in
encouraging collaborators to attend. Training events can start with
generic material, shared between experiments, then move to separate
experiment specific topics. This would be especially useful for
experiments sharing a home laboratory and provides an additional benefit
when different groups have managed to converge on common components.

Training and documentation are intimately linked and good training
materials can be studied offline by students in their own time.

As much use as possible should be made of generic resources that already
exist; there is a role here for the HSF to curate useful training
materials (where their ongoing usefulness and validity needs to be
managed) and encourage cooperation with other training initiatives, such
as Software Carpentry~\cite{SoftwareCarpentry}.

\hypertarget{seminars-and-discussion-forums}{%
\subsubsection{Seminars and discussion
forums}\label{seminars-and-discussion-forums}}

These are important to the community, in particular to discuss new
technologies or tools at an early stage of development. A good example
is the Software Technology
Forum~\cite{SoftwareTechForum}. These forums also allow questions to be posed
about problems that are encountered and identifying existing software that may provide
a solution (thus avoiding re-inventing the wheel yet again).

In terms of knowledge sharing among developers, solutions such
Stack Overflow are excellent for establishing
communication between software developers and publishing answers on
the web. In fact, this solution is used by many developers to find
solutions to the daily problems they face. It seems that in general
HEP software developers are consumers of these general help lists and we
encourage the community to take more active role in contributing as
well. Where a smaller community does not have the critical mass for
their own Stack Exchange service other products, like Discourse, can be
installed on-premises.

In addition, attempts have also been made to set up general software
discussion mailing lists in the past. However, it seems that after some
initial flurry of activity these are not really very successful. They
may be too far away from most developers day-to-day concerns and there
is probably some reluctance to pose questions to a wide list. Though
being a fast and attractive solution for discussions about a particular
issues they are not as viable for long term or broad knowledge sharing.
There may be some use in trying alternatives, such as a Slack channel,
at least to prototype with low investment.

\hypertarget{workshops-and-community-reviews}{%
\subsubsection{Workshops and community
reviews}\label{workshops-and-community-reviews}}

Hosting workshops on specific topics is a particularly useful way to
bring together experts from different experiments. These have an
essential role in introducing new ideas to the community and providing
an environment in which people can meet and interact. The HSF has also
organised community meetings on aspects of future HEP software
development that are of particular interest to the community in a
strategic sense. Two such meetings have been organised so far: one on
Geant V simulation~\cite{HSF-GeantV-Report}, and the other on the analysis
ecosystem~\cite{HSF-Analysis-Report}. These meetings can provide very valuable
feedback to developers on the priorities of the experiments and the most
critical features to develop and support. Organising further meetings on such topics provides
important strategic feedback on development.

\hypertarget{conferences}{%
\subsubsection{Conferences}\label{conferences}}

HEP has a strong tradition of software and computing conferences (ACAT
and CHEP are good examples). These continue to be a critical part of the
lifecycle of software and of the community that supports it, offering
recognition of work done as well as dissemination of knowledge about
available solutions that can, hopefully, be adopted by others. Although
they focus on completed pieces of work, they complement other activities
well and provide an important space for social interaction.

It would be beneficial for the field as a whole to increase its
participation in general conferences and software focused events. This
would help generate relationships and collaborations outside the direct
field with experts who may have interesting observations and
suggestions.

\hypertarget{journal-publication}{%
\subsubsection{Journal publication}\label{journal-publication}}

Journals disseminate information to the wider community in a
permanent way and so form an important part of the collective memory of the
community. Journals also have another significant role to play
in recognition of software work, which is discussed at greater length in
Section~\ref{recognition-of-software-development}.

\hypertarget{reusing-code-effectively}{%
\subsection{Reusing code effectively}\label{reusing-code-effectively}}

Effective reuse of code requires good design, documentation and support.
Good design starts from a good understanding of the problem to be
solved, with an appreciation of the important external interface to
provide to clients. APIs should be stable, with scope to change the internal
details of an implementation as needed. Training software developers in
good design principles will certainly help in this regard; however,
often the reuse potential of code is only discovered post-facto and then some
advice and expertise in refactoring will be useful for the longevity of
the package. Here Agile methodologies can help to focus on the key areas
of responsibility and avoid over designing interfaces for imagined
situations.

There is scope to improve the training in these areas within the HEP
community as discussed above.

In all cases good documentation for code will greatly help adoption (the
same solutions apply here as were discussed above) and this provides a
key aspect of user support. Utilising popular standard development
tools encourages collaboration and provides issue tracking capabilities.

\hypertarget{software-licenses}{%
\subsubsection{Software Licenses}\label{software-licenses}}

All code should have a clear copyright holder, which gives the legal
ownership of the software. Applying a license specifies how the software
can be used by others. Open-source licenses are now widely accepted by
the HEP experiments, which we believe is the correct approach. Open
source licenses fall into various categories

\begin{itemize}
\item
    Copyleft licenses mandate that any public releases with changes to the
  code must be licensed under the same terms and, in addition, that any
  software which uses the copyleft code must also itself be distributed
  under a copyleft license (which is why these licenses are sometimes
  referred to as `viral'). The General Public License, GPL, is the best
  known of these.
  \item
    Weak copyleft licenses (such as the Lesser General Public License,
  LGPL or the Mozilla Public License, MPL) remove the latter copyleft
  via use requirement, but still mandate that direct modifications to
  the package itself be redistributed under the same terms.
  \item
    Permissive licenses (Apache, MIT, BSD) generally allow for public
  modifications to be made without requiring them to be released under
  the same terms as the original code.
\end{itemize}

There is a modern report from the HSF on licensing 
topics~\cite{HSF-TN-2016-01} and we do not
repeat the discussion here. Instead we encourage code authors to ensure
that copyright and license for code is well established at the beginning
of the project, including the status of contributions to the code. The
implications of the chosen license should be properly understood,
especially in so far as it affects contributors to the code base and
users of the code who may incorporate it into a larger software stack,
together with their own code (the GPL has a fairly profound effect in
this regard).

\hypertarget{validation-of-complex-outputs}{%
\subsection{Validation of complex
outputs}\label{validation-of-complex-outputs}}

The outputs of HEP software packages can be extremely complex and thus
difficult to validate. This is especially true of the simulation of
modern high energy physics detectors, which are incredibly complex
machines and need to be accurately modeled to understand their
performance for rare decays and weak signals. Up to now most experiments
have developed their own infrastructure for comparing results.
Comparison of distributions is a relatively well understood topic in
statistics, so having standard tools that help compare outputs from
different versions of software would help save effort between
experiments and offers scope for improving the quality of our
validation.

\hypertarget{data-and-code-preservation}{%
\subsection{Data and code
preservation}\label{data-and-code-preservation}}

The longevity of code in HEP is becoming increasingly important. Large
experiments have multi-decade active lifetimes, with ``data
preservation'' now considered an important topic after an experiment has
finished taking data. Data taken in the early years of the experiment
may need to be reanalysed years afterwards with older software that has
not been built for a long time. Software virtual machines or containers
can help a great deal in preserving the working environment, including
the source code itself. Should changes need to be made to software to
correct for a long dormant bug or add a necessary new feature, having a
virtual machine or container with the source code and build environment,
in addition to any binary products, is important. Here the use of
standard tools, as encouraged above, can help, with the state of the
code and externals being encapsulated in a source repository tag. This
is becoming a prerequisite to publication in some journals, so being
able to trace the software used to produce some results, all the way
through to an analysis, is crucial. Experiments need to strongly
encourage their members to work in a preservable way.

In the design of code, relying on standard language features is more
sustainable than utilising external libraries that may become
deprecated. Of course, libraries that do tasks well are preferable to
inferior home grown implementations and achieving some uniformity in the
community will help for longer term support issues when they arise.

For further details on this topic we refer to the Data Preservation Community
White Paper group~\cite{HSF-CWP-2017-06}.

\hypertarget{code-deprecation}{%
\subsection{Code deprecation}\label{code-deprecation}}

The complexity of HEP software and the generic nature of many of the
sub-tasks that need to be performed mean that many external libraries
are essential to our workflows. It would be completely unfeasible and a
waste of effort to reproduce such code internally. However, there are
often many solutions to the same problem and picking the one that has
the best long term prospects or becomes the most widely adopted is not
possible a priori. Sometimes externals will simply be eclipsed by
alternatives and sometimes they will change their APIs in
backward-incompatible ways. The problem is often exacerbated by the fact
that the HEP-specific client may itself be only in maintenance mode and
the original author may not even work in the field anymore.~

In these cases evolving away from deprecated code is required. This can
be a labor intensive operation, but often many aspects of it are
automatable. There is a good example in the refactoring of C++, which is
a hard problem, but with the development of Clang and full knowledge of
the abstract syntax tree (AST) of the code, there are significant
possibilities for using this knowledge to tackle much of the drudge work
of refactoring. There have been successful examples of using Clang in
this way~\cite{ClangMR41342}
and we believe this is an area where investment could be very profitable,
both specifically for HEP and as a contribution to the wider open-source
community. Tools to further help validate the updated code can be shared
between experiments, e.g., support for any migration to new versions of
ROOT~\cite{Brun1996} or Geant4~\cite{Agostinelli2003}.

Changing software versions and evolving platforms is a natural part of
the lifecycle of production environments and can be helped by migration
to more self-contained deployment models, e.g., by using containers
instead of a site provided environment.

\hypertarget{hardware-and-software-evolution}{%
\subsection{Hardware and software
evolution}\label{hardware-and-software-evolution}}

The evolution of processors in recent years have given rise to highly
parallel architectures (from many core CPUs to GPUs); in addition,
thanks to the push coming from the mobile market, viable low-power
solutions are also now common (e.g., ARM architectures).

\hypertarget{effective-exploitation-of-parallel-hardware}{%
\subsubsection{Effective exploitation of parallel
hardware}\label{effective-exploitation-of-parallel-hardware}}

Implementation and optimisation of software on parallel architectures
requires different skills, tools and training (generally more advanced)
than the ones required for the serial world. In order to allow
physicists to effectively exploit parallel architecture, means are
required to insulate end user programmers from many of the low-level
difficult details of parallel programming.

Many specific details of parallel architectures can be masked by the
adoption of appropriate libraries. Generally HEP problems fit better
into a task based scheduling pattern, rather than to an optimiser of
kernels run in tight loops. The community has utilised Intel's Threaded
Building Blocks (TBB)~\cite{Pheatt:2008:ITB:1352079.1352134,TBB} successfully in
many areas. Should TBB continue to develop and be a good fit for our problems
this common solution also helps solve the problem of resource balancing between different parts of
the code. Should the solutions adopted become fragmented then overall
resource balancing becomes much harder and may become difficult to
solve. Evolution of the C++ standard will most likely bring more
concurrency primitives into the language, however, so some evolution
here is to be expected and should be planned for. This would mean
appropriate zero cost abstractions when possible.

\hypertarget{achieving-performance-portability}{%
\subsubsection{Achieving performance
portability~}\label{achieving-performance-portability}}

The fact that different parallel architectures (e.g., CPUs and GPUs) have
different sweet-spots for performance makes compile time performance
portability extremely difficult to obtain. This is true across
architectures, but also between different generations of the same
architecture. Past abstraction techniques might not work and development
needs to target specific hardware for best performance. However, it is
strongly advisable that the design of the code separates physics logic
from performance optimisation. This helps to avoid premature
optimisation and a harmful reduction in code maintainability. Options
for runtime optimisation would also be possible (e.g., fat libraries
with code for various vector architectures) and may help to avoid an
explosion in build target architectures. Such techniques are an area
where common community investigation would be worthwhile, especially
when common tools are used in multiple projects.

When aiming for cross-architecture performance portability, one
important concern is data layout, as different architectures have
different preferences in this respect. For example, GPUs perform best
when memory accesses from different threads are coalesced (as it
optimises memory bus utilisation), whereas CPUs perform best when writes
from different threads are relatively far apart in RAM (to avoid the
cache trashing effect known as ``false sharing''). These differences can
be abstracted away through the use of data structure generation mechanisms,
which are still in their early days but are begining to be offered by some
programming environments such as
Kokkos~\cite{CarterEdwards20143202},
and some domain specific languages, such as those used in the context of the
H.E.S.S.\ Observatory~\cite{HESS} and in the Cherenkov Telescope Array
reconstructions~\cite{CTA}.

\hypertarget{tracking-of-new-software-technologies-and-tools}{%
\subsubsection{Tracking of new software technologies and
tools}\label{tracking-of-new-software-technologies-and-tools}}

Compared to when HEP started as a field, the importance and ubiquity of
software development has increased hugely. If in the late 80s and 90s
many of the tools and practices were actually started inside our field
itself, or at least within academia, it is now true that a large amount
of development happens outside the traditional HEP field. Starting from
the contributions of ``bedroom coders'', to the scale of large
corporations like Google, Apple, or Facebook, a large number of high
quality tools and libraries are now available as open-source projects
for everyone to use. While this affords the field great opportunities,
the rate at which these tools and libraries change is something that
necessitates a continuous attention and an in-depth scrutiny of their
potential, as well as managing inevitable deprecations. As was
pointed out above, pooling community knowledge here will make optimal
choices in new technology more likely.

\hypertarget{dealing-with-hardware-diversity}{%
\subsubsection{Dealing with hardware
diversity}\label{dealing-with-hardware-diversity}}

The increasing number of different architectures, operating systems and
compilers with different versions give rise to a large number of
possible platforms that combine these parameters. Experiment software
should be tested on a large range of platforms to ensure a
competitive and effective use of these new technologies, which requires
good test cases and striving to minimise the operational and human costs
of multiple platforms. Use of open-source orchestration toolkits for
managing these stacks can help a lot. Jenkins has received much
attention, but alternatives, such as Travis, exist and there may be scope
for a community evaluation. Aspects such as application independence,
adaptation and modularity (a plugin-based structure) are important.
Basic platform types can share a lot of common infrastructure (e.g.,
x86\_64, AArch64 and PowerPC) that minimise the cost of adding new
flavors if common components are used.

There is an important link here when deciding what level of validation
is considered acceptable on each platform, especially if platforms do
not produce binary identical results. In order to evaluate and prepare
for the incorporation of such new platforms having early access is
extremely important. CERN's TechLab and OpenLab projects help provide
that, but there is a need to ensure that the whole community can benefit
and more sites could be involved in this activity.

\hypertarget{software-deployment}{%
\subsection{Software deployment}\label{software-deployment}}

Deployment of HEP software has been revolutionised by the widespread
adoption of CERN Virtual Machine File System (CVMFS)~\cite{7310920}
for flexible and ubiquitous access to a common
filesystem namespace with efficient distribution of content that can be
strongly cached at multiple levels. We expect that CVMFS will remain the
mainstay of software distribution to most computing sites in the
foreseeable future.

However, CVMFS cannot cover all possible use cases. Sites that have no
outbound network access (common in the HPC world) or are mobile and
perhaps temporarily disconnected (e.g., developer laptops) require
additional distribution mechanisms. Here, containerisation offers a
great deal of promise as a way to isolate the software environment from
the underlying hardware and to conveniently provide all the needed
runtime components. This is particularly useful when other constraints
drive the base operating system away from the standard one adopted and
validated by the experiment. At the moment Docker is favored as being
exceptionally convenient for development, with automated conversion to
other container implementations (e.g., Shifter or Singularity) being
possible. There should be no reason to deeply bind anything in the
install process into any particular container implementation.

Containers also play a role in data and code preservation as already
noted. While containers are architecture specific, there is currently no
serious risk that, e.g., x86\_64 platforms will be unavailable in the
foreseeable future.

\hypertarget{recognition-of-software-development}{%
\subsection{Recognition of software
development}\label{recognition-of-software-development}}

Proper recognition of the contribution that software makes to the
success of High Energy Physics is a key element of sustaining that
effort and establishing software as a `first class citizen' in the
field. As an academic discipline, HEP primarily recognises publications
in refereed academic journals and the citations of those publications as
indicators of merit. Therefore we strongly encourage authors of software
to submit publications of that software to such journals, of which there
are now many~\cite{SSI2017}.
It is also equally important that publications from the field
appropriately cite the software that they have used.

The correct level of granularity of software versions for publication
will need to be judged by authors themselves, as well as reviewers.
Significant improvements in algorithms, in either machine resource
consumption or in physics performance, would certainly merit a new
publication, as might a major release of software with significant new
features. There is also a strong case for having an overall publication
that encompasses a major piece of combined software used by an
experiment. One example is an article on an experiment's offline
software stack (that itself includes citations of the other software
used within the combined software). Such publications would then be
highly citable in physics papers, analogous to the common citations of
the detector design papers.

Currently, if HEP software has an article written about it, it will tend
to be in the conference proceedings of, e.g., CHEP or ACAT. Although
these are refereed, in general the barrier for entry is low, and as such
they are not considered as prestigious as refereed journals as the
quality is mixed. Without some reorganisation of the community, and a
more strict refereeing process, it is unlikely that conference
proceedings will become the journals of choice for recognition in our
field. However, conference journals (e.g., Journal of Physics:
Conference Series) usually do not allow resubmission of articles to
other journals without extended content. Software authors should be
aware of these issues when deciding where best to publish.

There are also now interesting projects to directly publish software
itself, for example to Zenodo~\cite{Zenodo} or
figshare\cite{figshare}, as has been recommended by the
FORCE11
Software Citation Working Group~\cite{10.7717/peerj-cs.86}. We are very
encouraging of such efforts, and believe it would be useful for HEP
software authors to participate, but at the moment the actual
recognition accrued from such a publication is unclear. Through its work
in reviewing papers, the HEP community could choose to suggest to
authors that they cite the software they use, whether directly or via
software papers. Over time, this small effort on the part of reviewers
could be sufficient to change community practice regarding software
credit.

For further discussion on this topic see also the Careers and
Training Community White Paper~\cite{HSF-CWP-2017-02}.

\hypertarget{roadmap-for-the-next-10-years}{%
\section{Roadmap for the next 10
years}\label{roadmap-for-the-next-10-years}}

In the areas covered by this white paper HEP must endeavor to be as
responsive as possible to developments outside of our field. In terms of
hardware and software tools there remains great uncertainty as to which
platforms will offer the best value for money in a decade. It therefore
behooves us to be as generic as possible in our technology choices, retaining
the necessary agility to adapt to this uncertain future.

Our vision is characterised by HEP being current with technologies and
para\-digms that are dominant in the wider software development world,
especially for open-source software, which we believe to be the right
model for our community. In order to achieve that aim we propose that
the community establishes a development forum that allows for technology
tracking and discussion of new opportunities. The HSF can play a key
role in marshalling this group and in ensuring its findings are widely
disseminated. In addition, having wider and more accessible training for
developers in the field, that will teach the core skills needed for
effective software development, would be of great benefit.

Given our agile focus, it is better to propose here projects and
objectives to be investigated in the short to medium term, alongside
establishing the means to continually review and refocus the community
on the most promising areas. The main idea is to investigate new tools
as demonstrator projects where clear metrics for success in reasonable
time should be established to avoid wasting community effort on
initially promising products that fail to live up to expectations.

\hypertarget{short-term-projects-1-2-years-and-ongoing-activities}{%
\subsection{Short term projects (1-2 years and ongoing
activities)}\label{short-term-projects-1-2-years-and-ongoing-activities}}

\begin{itemize}
\item
    Establish a common forum for the discussion of HEP software problems.
  This should be modeled along the lines of the Concurrency
  Forum~\cite{ConcurrencyForum}, which was very successful in establishing
  demonstrators and prototypes that were used as experiments started to develop
  multi-threading frameworks.
  \item
  Continue the HSF working group on packaging, with more prototype
  implementations based on the stronger candidates identified so far.
  \item
    Provide practical advice on how to best set up new software packages,
  developing on the current project template work and working to
  advertise this within the community.
  \item
    Work with HEP experiments and other training projects to provide
  accessible core skills training to the community. This training should
  be experiment neutral, but could be usefully combined with the current
  experiment specific training. Specifically this work can build on and
  collaborate with recent highly successful initiatives like the LHCb
  StarterKit~\cite{1742-6596-898-8-082054} and ALICE Juniors~\cite{ALICEJuniors}
  and with established generic training initiatives such as Software
  Carpentry~\cite{SoftwareCarpentry}.
  \item
    Strengthen links with software communities and conferences outside of
  the HEP domain, presenting papers on the HEP experience and problem
  domain. SciPy, SC (informally known as Supercomputing), RSE
  Conference, and Workshop on Sustainable Software for Science: Practice
  and Experiences (WSSSPE) would all be useful conferences to consider.
  \item
    Write a paper that looks at case studies of successful and
  unsuccessful HEP software developments and draws specific conclusions
  and advice for future projects.
\end{itemize}

\hypertarget{medium-term-projects-3-5-years}{%
\subsection{Medium term projects (3-5
years)}\label{medium-term-projects-3-5-years}}

\begin{itemize}
\item
    Prototype C++ refactoring tools, with specific use cases in migrating
  HEP code.
  \item
    Prototyping of portable solutions for exploiting modern vector
  hardware on heterogenous platforms.
  \item
    Develop tooling and instrumentation to measure software performance,
  especially in the domain of concurrency. This should primarily aim to
  further the developments of existing tools, e.g.,
  igprof~\cite{Eulisse:2005zz}, rather than developing new ones.
  \item
  Develop a common infrastructure to gather and analyse data about
  experiments' software, including profiling information and code
  metrics, to ease sharing of this information across experiments.
  \item 
  Develop tool(s) that allow developers to understand the usage of their
  software and to be credited for this usage, possibly via using these metrics
  associated with software citation.
  \item
    Undertake a feasibility study of a common toolkit for statistical
  analysis that would be of use in regression testing for experiment's
  simulation and reconstruction software.
\end{itemize}

It would be highly beneficial for the group to continue to function,
through the means of the forum proposed above, and to have at least one
meeting a year devoted to reviewing our activities and updating our
knowledge of best tools and practices.

\sloppy
\raggedright
\clearpage
\printbibliography[title={References},heading=bibintoc]

\end{document}